\documentclass[
    ,final            
  ]
  {aipproc}

\layoutstyle{6x9}

\begin{document}

\title{GALEX measurements of the Big Blue Bump as a tool to study bolometric corrections in AGNs}


\author{Elena Marchese, R. Della Ceca, A. Caccianiga, A. Corral and P. Severgnini}{
  address={INAF-Osservatorio Astronomico di Brera, via Brera 28 - 20121 Milano, Italy\\
 e-mail:elena.marchese@brera.inaf.it}
}

\begin{abstract}
 Active Galactic Nuclei emit over the entire electromagnetic spectrum with the peak of the accretion disk emission in the far-UV, a wavelength range historically difficult to investigate. We use here the GALEX (Galaxy Evolution Explorer) Near-UV and Far-UV measurements (complemented with optical data from Sloan Digital Sky Survey (SDSS) and XMM-Newton X-ray spectra) of a sample of  83 X-ray selected type 1 AGN extracted from the XMM-Newton Bright Serendipitous Survey (XBS) to study their spectral energy distribution (SED) in the optical, Near and Far-UV and X-ray energy bands. We have constrained the luminosity of the accretion disk emission component and calculated the hard X-ray bolometric corrections for a significant sample of AGN spanning a large range in properties (z, $L_x$). \\

\end{abstract}

\maketitle


\subsection{Data Sources}The starting point of our study is a sample of  267 type 1 AGNs belonging to the XBS (for full details see \cite{DellaCeca} and \cite{Caccianiga}). By cross-correlating them with GALEX (http://www.galex.caltech.edu), a NASA Small Explorer mission covering the wavelength range from 1344  	\AA \ to  2831   \AA , and SDSS (http://www.sdss.org) data, we selected 83 Type 1 AGN, having at least one measured flux in the UV (FUV or NUV),  optical magnitudes from the SDSS and X-ray spectra from XMM-Newton .

\subsection{Corrections to fluxes and Hard X-ray bolometric corrections }

The data points from the SDSS and from GALEX were described using a basic accretion disk model (DISKPN model in the XSPEC package); the disc temperature was fixed to a typical value of 0.003 keV, while the normalization of the accretion disk was set to reproduce the GALEX NUV and FUV data. To derive the continuum emission the observed fluxes were corrected for the contribution of the strongest emission lines (Ly $\alpha$,  NV, CIV, MgII) and for  Galactic extinction, using  the law formulated by Allen (\cite{Allen}) and the galactic $E_{B-V}$ along the line of sight; we also corrected for the intrinsic extinction at the source redshift by using the extinction curve derived by Gaskell \& Benker  (\cite{Gaskell}).  The intrinsic source reddening, $E_{B-V}$, has been evaluated using the Galactic gas to dust ratio and the absorption column density (or upper limits) derived directly from the best fit X-ray spectral model. Finally we have estimated the AGN bolometric luminosity by adding to the intrinsic accretion disc luminosity (i.e. that corrected both for Galactic and intrinsic extinction)  the 0.1-100 keV luminosity, which was obtained, for each AGN, by extrapolating its properties ($L_x$, spectral index), as derived from the XMM-Netwon spectral data (Corral et al. 2009, this proceeding).\\
\textbf{  Preliminary results - }The preliminary results of this study are: \\
\textbf{a)} a very large spread in the distributions of the  hard X-ray bolometric corrections  ( from 5 up to 100 ), probably implying a large dispersion in the mean SED. There is no clear correlation between the bolometric correction and the X-ray luminosity (Fig.1 left) ;\\
\textbf{b)} a correlation between the bolometric correction and the photon index, probably related to the source accretion rate (Fig.1 right). \\
Further work/analysis is in progress.\\
\begin{figure}

\includegraphics[height=.295\textheight]{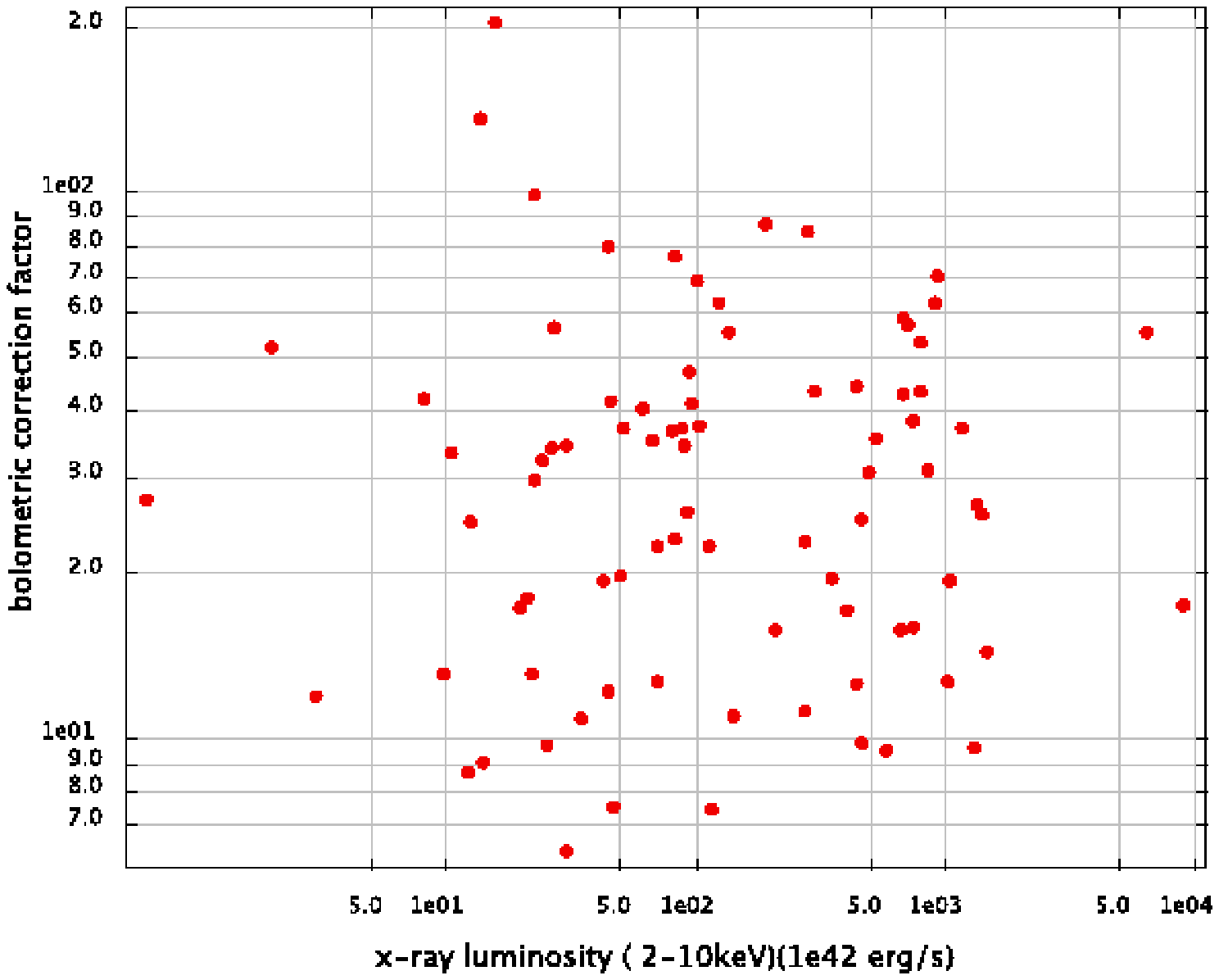}
\includegraphics[height=.295\textheight]{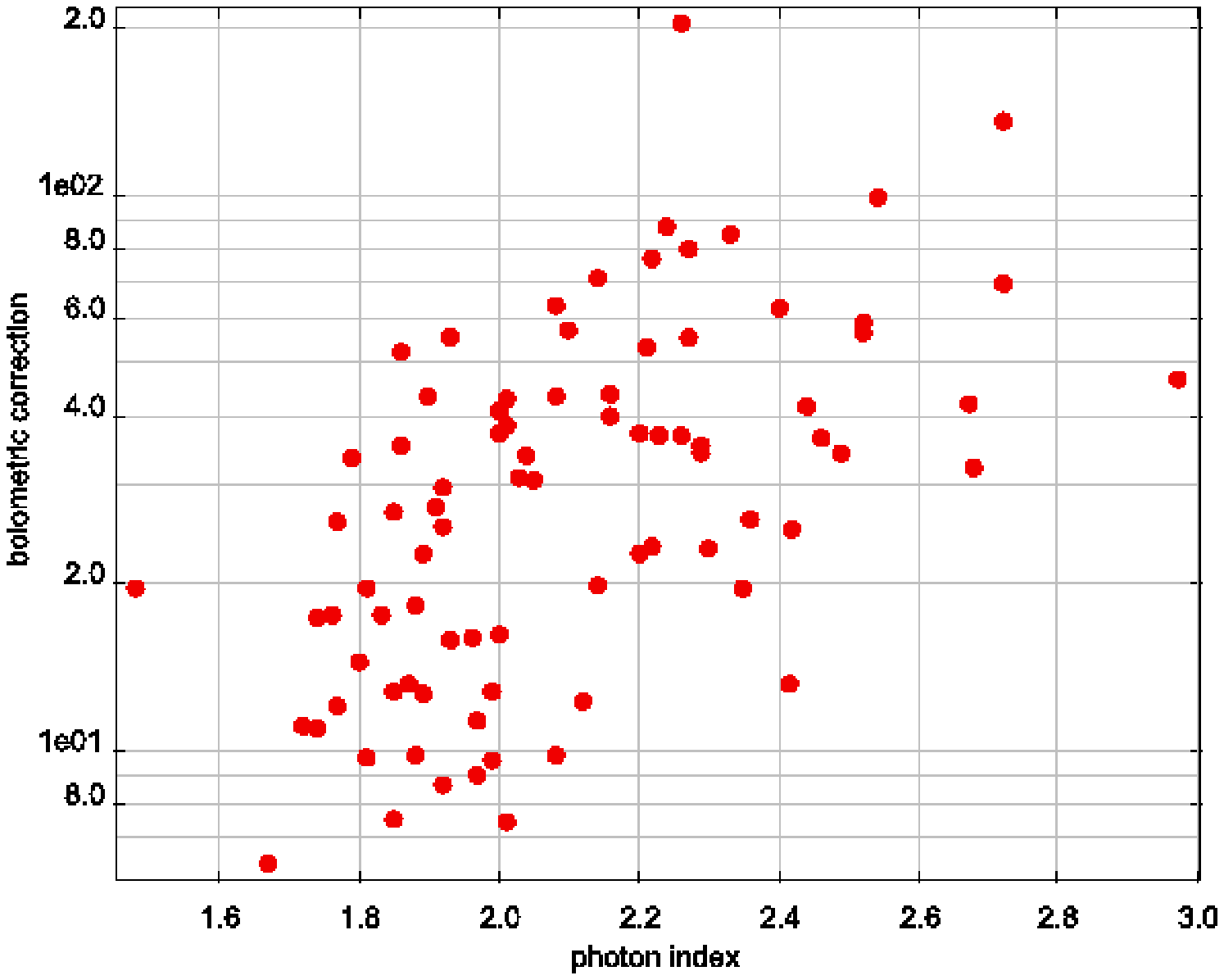}
\caption{  \emph{Left}: Bolometric correction  ($L_{bol}/L_{2-10 keV}$) as a function of $L_{2-10 keV}$.     \emph{Right}:Bolometric correction against the measured photon index; both quantities are probably related to the source accretion rate, see e.g. \cite{Vasudevan}  for a correlation between accretion rate and bolometric correction  and \cite{Atlee}   for a correlation between the photon index and the bolometric correction. }
\end{figure}

\begin{theacknowledgments}
   We acknowledge financial support from ASI (grant n.I/088/06/0 and COFIS contract).
\end{theacknowledgments}

\bibliographystyle{aipproc}   


\IfFileExists{\jobname.bbl}{}
 {\typeout{}
  \typeout{******************************************}
  \typeout{** Please run "bibtex \jobname" to optain}
  \typeout{** the bibliography and then re-run LaTeX}
  \typeout{** twice to fix the references!}
  \typeout{******************************************}
  \typeout{}
 }

\end{document}